\documentclass[preprint,12pt]{elsarticle}
\biboptions{numbers,sort&compress}

\usepackage{amssymb}
\usepackage{upgreek}
\usepackage{xcolor}

\usepackage{amsmath}

\journal{arXiv}

\begin{document}

\begin{frontmatter}



\title{Deposit of Red Blood Cells at low concentrations in evaporating droplets: central edge growth and potential applications}


\author{Vahideh Sardari$^{a,b}$}

     \author{Mahsa Mohammadian$^b$}
   
     \author{Shima Asfia$^b$}
      
      \author{Felix Maurer$^b$}

      \author{Diana \"Or\"um$^b$}

     \author{Ralf Seemann$^b$}
     
	\author{Thomas John$^b$}

    \author{Lars Kaestner$^{b,c}$}

     \author{Christian Wagner$^{b,d}$}
   
     \author{Maniya Maleki$^a$}

     \author{Alexis Darras$^b$}  
     \address{$^a${Department of Physics, Institute for Advanced Studies in Basic Sciences (IASBS), Zanjan, 45137-66731, Iran}

}

\address{$^b${Department of Experimental Physics \& Center for Biophysics, Saarland University, Saarbruecken, D-66123, Germany}

}

\address{$^c${Department of Theoretical Medicine and Biosciences, Saarland University, Homburg, D-66421, Germany}
}

\address{$^d${Physics and Materials Science Research Unit, University of Luxembourg, Luxembourg, L-4365, Luxembourg}
}

\begin{abstract}
Evaporation of blood droplets and diluted blood samples is a topic of intensive research, as it is seen as a possible low-cost tool for diagnosis. So far, samples with volume fraction down to a few percents of Red Blood Cells (RBCs) have been studied, and those were reportedly dominated by a ``coffee-ring'' deposit. In this study, samples with lower volume fractions have been used in order to study the growth of the evaporative deposit from sessile droplets more in details. We observed that blood samples and salt solutions with less than 1\% volume fraction of RBCs are dominated by a central deposit. We characterized the growth process of this central deposit by evaporating elongated drops, and determined that it is consistent with the Kardar-Parisi-Zhang process in the presence of quenched disorder. Our results showed a sensitivity of this deposit size to the fibrinogen concentration and shape of the RBCs, meaning that this parameter could be used to develop a new and cost-effective clinical marker for inflammation and RBC deformation.
\end{abstract}

%
%
%
%
%

\end{frontmatter}


\section{Introduction}
\label{Intro}

The formation of a dried stain of blood is a phenomenon that has been the focus of a lot of research interest, with applications ranging from low-cost diagnostic methods, e.g. for turbeculose, thalassemia and neonatal jaundice \cite{sikarwar2016automatic,bahmani2017study}, to forensic investigations \cite{liu2020automatic,attinger2022using}. In the soft matter community, the evaporation of colloidal droplets resting on a substrate, i.e., a sessile droplet, has been an area of intensive research in recent decades, since the seminal work of Deegan et al. \cite{deegan1997capillary}. Various mechanisms have been highlighted as of fundamental importance for the final structure of the deposits, an omnipresent one being the coffee ring effect. This coffee ring effect names the fact that an outward flow is usually dragging the particles near the pinned edge of the droplet, due to the inhomogeneous evaporation rate along the surface of the droplet \cite{deegan1997capillary}. However, when a high enough salt concentration is considered, the dominating mechanism is a Marangoni recirculation, due to the gradient of salt concentration. Indeed, an important increase of salt concentration also leads to an increase in surface tension, and gradient of concentrations therefore can induce a significant gradient of surface tension, which can create a central deposits of the particles \cite{rossi2019interfacial,darras2018transitional,darras2019combined}.\\ 
In the case of blood evaporative deposits, the roles of temperature \cite{pal2021temperature}, of the ambient humidity \cite{zeid2013influence}, of the substrate’s nature, and of ions and proteins present in the plasma \cite{chen2016blood,pal2023drying} on the final structure have been extensively studied. Moreover, seminal studies showed an effect of Red Blood Cell (RBC) properties by comparing rigidified cells or spheres with healthy RBCs \cite{lanotte2017role}, while others have highlighted the effect of dilution by studying patterns obtained from diluted full blood in distilled water and Phosphate Buffered Saline (PBS), down to full blood concentration of around $10\%$ \cite{pal2020concentration}. However, the detailed influence of RBC properties, such as shapes and aggregability, on the deposit growth and final state is still unclear. This is nonetheless particularly relevant, since previous studies evidenced that, in the case of solid particles, anisotropy of the colloids could revert the deposit pattern by itself \cite{yunker2011suppression}, or dictate their growth dynamics \cite{yunker2013effects}. \\
The main objective of this study is to determine whether evaporative deposits of dilute RBC suspensions could be used as low-cost screening methods for pathological modifications of RBC aggregability and shapes. Therefore, we focus on droplets of with volume fraction $\phi<1\%$ of RBCs (compared to the physiological level of $\phi\approx 45\%$) in plasma- and Phosphate Buffered Saline (PBS)-based solutions. We observed a central deposit, for the various suspending liquids with isotonic osmolarities. Exploiting this result, we create line-shaped droplets and study how the cell shape and aggregation modify the growth and final central linear deposit. We mixed plasma with serum to modify the aggregation of RBCs \cite{dasanna2022erythrocyte}, while osmolarity of the PBS-based suspensions was varied to modify the shape of the RBCS. All conditions led to a growth consistent with the Kardar-Parisi-Zhang process in the presence of quenched disorder (KPZQ) \cite{yunker2013effects,csahok1993dynamics}. However, the final sizes of the deposits vary as a function of the cell shape, volume and aggregability, meaning that the study of those patterns could potentially be used as a low-cost screening indicators for deformed RBCs or abnormal aggregation levels.\\

\section{Methods}

\subsection{Ethics statement}
Human blood withdrawal from healthy volunteers was performed after explicitly obtaining their informed consent. Blood withdrawal and handling were performed according to the declaration of Helsinki and the approval by the ethics committee "Aerztekammer des Saarlandes" (reference No 176/21). \\

\subsection{Experimental protocol}
Fresh blood was drawn from healthy donors via venous blood sampling in tubes containing Ethylenediamine tetraacetic acid (EDTA) and a tube for serum (S-Monovette, Sarstedt, Nuembrecht, Germany). Cells were separated from the plasma by centrifuging the EDTA tubes for $7$ minutes at $3000\,\mathrm{rcf}$ using a fixed angle centrifuge (Hermle Z 36 HK,Hermle, Wehingen, Germany). Plasma is then obtained as the supernatant, and is transferred immediately after centrifugation into a clean 1.5 mL Eppendorf tube using a micro-pipette. Serum is collected in a similar fashion, except that the blood is permitted to clot by leaving the collection tube undisturbed at room temperature for 30 minutes, according to the serum-tube manufacturer's guidelines (S-Monovette, Sarstedt, Nuembrecht, Germany). Clotted components are afterwards removed by centrifuging with the same protocol as for the plasma. Serum is obtained as the supernatant. Afterwards, $1\,\mathrm{mL}$ samples with volume fraction $\phi=3\times 10^{-3}$ of RBCs were prepared by suspending the packed cells in PBS $(Gibco, USA)$, plasma and/or serum. Samples were kept at room temperature for the duration of the experiments (less than $6\,\mathrm{h}$).

In order to investigate whether changes in RBC shapes can lead to a different growth class of the edges obtained in the deposits, we also suspended RBCs in suspensions of various osmolarities. Indeed, in healthy individuals, RBCs at rest typically possess a biconcave disk shape. However, the osmolarity of the surrounding medium can change the shape of the RBCs according to the stomatocyte-discocyte-echinocyte (SDE) sequence \cite{brecher1972present,geekiyanage2019coarse,simionato2021red}. To that end, solutions with osmolarities of $(130, 200, 300, 550, 800 \,\mathrm{ mOsmol/L})$, with 5 mg/mL of Fibrinogen to ensure aggregation between RBCs, were prepared according to the protocols described in the next paragraphs. Washed RBCs were finally resuspended in each solution with the same volume fraction $\phi=3\times 10^{-3}$ as earlier.

To prepare the RBC suspensions with osmolarity lower than physiological and PBS values ($\thickapprox 300\,\mathrm{mOsmol/L}$), we first diluted Fibrinogen from human plasma (Sigma Aldrich, Burlington, Massachusetts, United States) in a solution of physiological osmolarity. Indeed, for stability reasons, human fibrinogen plasma is usually stored as a powder containing around 60\% of mass of the protein, stabilized with 15\% sodium citrate and 25\% NaCl. We determined experimentally that one can reproduce an iso-osmotic solution of Fibrinogen in PBS by using 5\% of distilled water for each 1 mg/mL of fibrinogen (i.e., for each $1/0.6\approx 1.67\,\mathrm{mg/mL}$ of the powder). However, due to possible heterogeneity in the powder, the osmolarity of this solution is always checked with a freezing-point osmometer (Osmomat 3000basic, Gonotec, Berlin, Germany). Given that this first step might require some repetitions to achieve the desired concentration of fibrinogen at the right osmolarity, and that fibrinogen is faster to dissolve at $300\,$mOsmol/L than in solutions of lower osmolarity, we always first prepared an iso-osmotic solution with a concentration of Fibrinogen equal to $F_t*Osm_{iso}/Osm_t$. Here we have $F_t=5\,\mathrm{mg/mL}$ as the targeted concentration of Fibrinogen, $Osm_{iso}=300\,\mathrm{mOsmol/L}$ the osmolarity of the isotonic solution and $Osm_t<Osm_{iso}$ the targeted osmolarity. The isotonic solution is then diluted with distilled water with proportions of $Osm_t/Osm_{iso}$ of the isotonic fibrinogen solution and $1-Osm_t/Osm_{iso}$ of distilled water. Finally, $45\pm5\,\mathrm{mg/mL}$ of Bovine Serum Albumine were added to the solution. Red Blood Cells were then suspended in each solution according to the protocol described previously.

To prepare the RBC suspensions with osmolarity higher than physiological (and PBS) values ($550, 800 \,\mathrm{ mOsmol/L}$), isotonic solutions with $5\,\mathrm{mg/mL}$ of Fibrinogen and $45\,\mathrm{mg/mL}$ of BSA were prepared as previously. Afterwards, a mass $m=\Delta C\,V\,M/2$ of NaCl was added to the solution, where $\Delta C$ is the required increased in osmolarity, $V$ is the volume of the solution and $M=58.44\,\mathrm{g/mol}$ is the molar mass of NaCl.

Final osmolarity of the solutions were always checked with a freezing-point osmometer and only used if the deviation was lower than 10\% from the targeted value.

\begin{figure}[ht!]
	\begin{center}
		\includegraphics[width=1.0\textwidth]{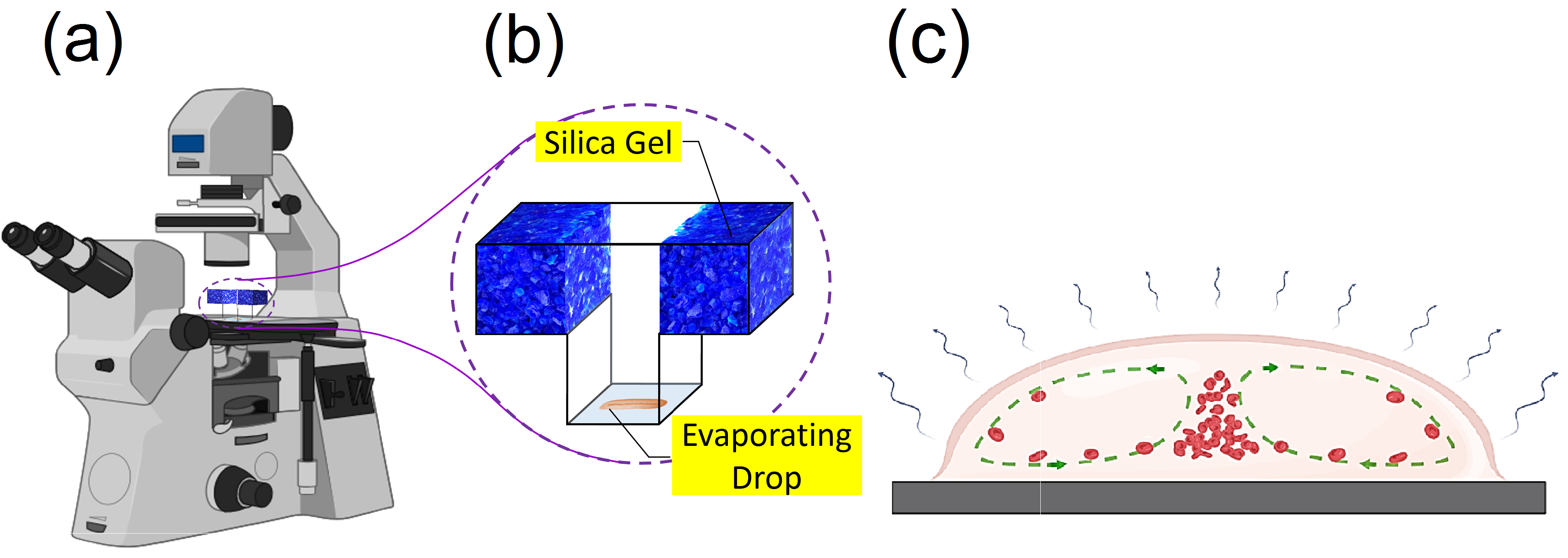}
		\caption{Experimental setup (a) schematic of inverted transmission microscope, and (b) the T-shaped drying chamber where the sample drop is placed on a microscopic slide. The slide contact to the chamber is sealed with petroleum jelly. (c) Schematic of the Marangoni flow inside the evaporating droplet, which transports the cells and gathers them in the center of the droplet. Such flow pattern is observed in solutions with high salt concentration \cite{rossi2019interfacial,marin2019solutal}.}
		\label{fig:figsetup}
	\end{center}
\end{figure}

\begin{figure}[ht!]
	\begin{center}
		\includegraphics[width=1.1\textwidth]{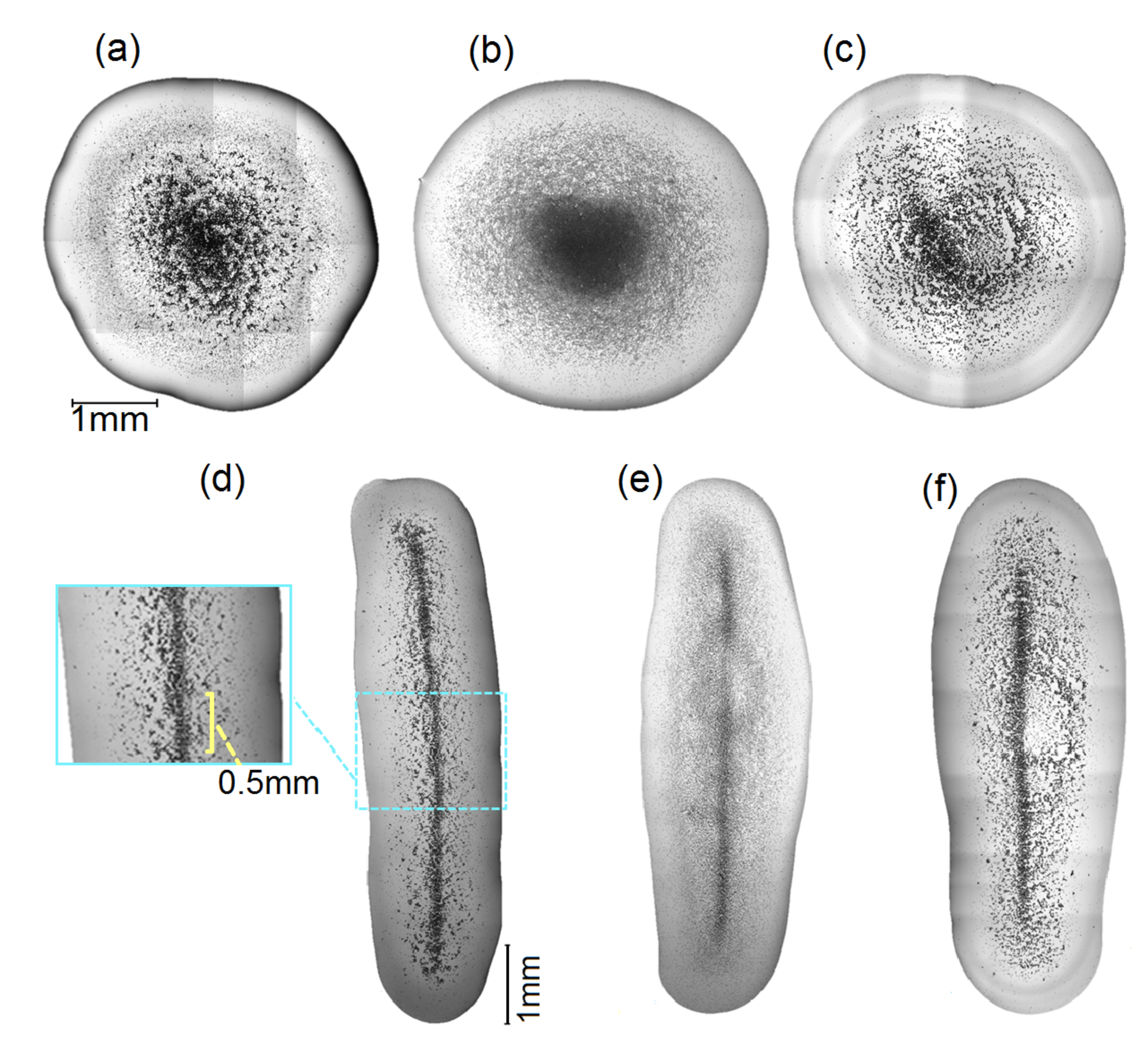}
		\caption{
			Pictures droplets of various suspensions of RBCs ($\phi=3\,10^{-3}$), shortly before complete evaporation and the apparition of cracks (see Supp Fig.S1 for completely dried deposits). (a-f) Whole drops were reconstructed by stitching series of pictures at higher zoom level. (a-c) Spherical cap droplets of $5\,\upmu\mathrm{L}$, (d-f): Elongated droplets of $5\,\mathrm{\upmu L}$ spread along $7\,\mathrm{mm}$ on the glass slide. (d) Insert shows the area of a single picture, where deposition of red blood cells was investigated along time. Supernatants were (a and d) autologous plasma, (b and e) $4.5\,\mathrm{mg/mL}$  BSA solution in PBS, and (c and f) $5\,\mathrm{mg/mL}$ fibrinogen solution in PBS with $4.5\,\mathrm{mg/mL}$ BSA. The pieces of images are stitched together with ImageJ.
		}
		\label{fig:fig1}
	\end{center}
\end{figure}

To create the deposits, a $5\,\mathrm{\upmu L}$ pendant drop of the sample was formed at the tip of an adjustable 2-20 $\upmu$L volume micropipette (Eppendorf, Hamburg, Germany). The pendant droplet was then placed on a glass slide (Micorscope cover glass, ECN 631-1586, VWR, Radnor, Pennsylvania), previously cleaned with isopropanol and dried with compressed air (contact angle at rest of PBS was measured to be $52\pm3^\circ$). To start an experiment with an elongated drolet, the $5\,\mathrm{\upmu L}$-drop was spread along 7 mm on the glass slide, using the same adjustable micropipette and a marked distance below the glass slide. To suppress the air flow and control the humidity of the environment in which the drop was placed for drying, a T-shaped hood was used, whose both top sides were filled with completely dry blue silica gels to absorb humidity, as described in earlier works \cite{darras2018remote,darras2019combined,darras2018transitional} (see Fig. \ref{fig:figsetup}). The glass slide with the droplet was placed under the microscope, and the T-shaped hood was adjusted on top of the glass slide with petroleum jelly (KORASILON-Paste, Kurt Obermeier GmbH, Bad Berleburg, Germany) on its base. The petroleum jelly ensures firm cohesion between the chamber and the glass slide, and prevents the passage of airflow into the chamber. The drop was observed under an inverted transmission microscope (Nikon Eclipse $TE200$, Nikon, Tokyo, Japan) by using a 4$\times$ objective. Illuminating white light sent through a short-pass filter with a cut-off wavelength of  450 nm (FESH0450, Thorlabs, Newton, New Jersay, USA). The filter was used in order to increase the contrast between the RBCs, whose absorption peak is at 415$\,\mathrm{nm}$ due to hemoglobin, and the surrounding environment. The images of the evaporating drop were taken at an interval of $1\,\mathrm{s}$ using a CMOS-camera (DMK 37BUX250, The Imaging Source, Charlotte, North Carolina, USA) attached to the microscope. Images are recorded until the droplet is completely dried and cracks appear (see Appendix Fig. \ref{fig:SuppDeposits}), i.e., around 40 min. Examples of the obtained pictures are depicted in Fig. \ref{fig:fig1}.

 \begin{figure}[ht!]
 	\begin{center}
 		\includegraphics[width=1.05\textwidth]{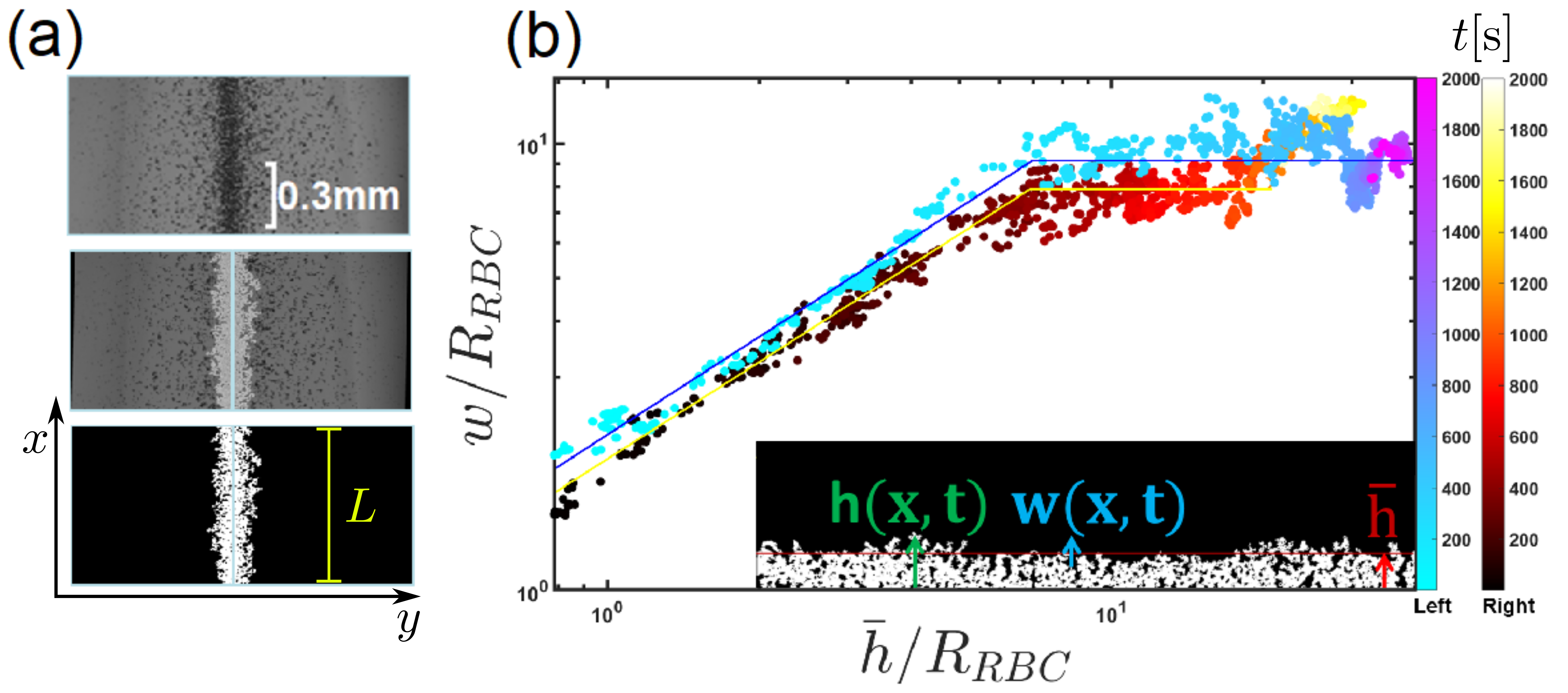}
 	
 		\caption{Illustration of the image analysis process. (a) Snapshot of the central deposit with the raw experimental picture split at the detected center (for the resulting pattern of $0.3\%$ volume suspension of RBCs in a 50-50 mixture of autologous serum and plasma), with overlaid final edges and the final binarized edges.  This picture was taken $35$ minutes after placing the drop on the slide. The $x,y$ axes and the length of the system $L$ are also indicated here for further references. (b) Scaling of the interface width with average height for superposition of the left side of the pattern and the right side of the pattern. Distances are normalized by the characteristic cell radius, in order to compare samples with possible radius changes (when various osmolarities are used). Solid lines are the fits used to obtain the parameter from Eq.(\ref{eq:BetaEq}). The slope of the first linear part, before saturation is reached, is the beta value.  Color bars indicate the elapsed time based on the frame number after placing the drop on the surface of the microscope slide.}
 		\label{fig:LEFTRIGHTFINAL}
 \end{center}
\end{figure}

 \subsection{Image analysis}
 Using a homemade Matlab code, the images of elongated droplets (see Fig. \ref{fig:fig1} (d-f)) were processed to binarize and isolate the central particle deposit. For each image, the code first averages the intensity perpendicularly to the main axis of the droplet, i.e., along $y$, and assesses the central position of the droplet as the $y$-coordinate with the minimal averaged intensity (since the RBCs have an absorption maximum at 420 nm, i.e., for blue light). The central deposit is then divided in two parts, splitting the picture at this central position, i.e., defining $y=0$, as illustrated in Fig. \ref{fig:LEFTRIGHTFINAL}(a). This effectively converts the central deposit as two growing edges. Analyzing both sides independently made it possible to monitor possible asymmetry in the droplet and/or deposit shape, which was used to reject data when significant differences were observed. Those pictures were then binarized using an Otsu threshold on pixel intensity. Afterwards, only the connected components in contact with the central edge at $y=0$ are considered as the growing edge, while other cells are discarded in further analysis. Furthermore, in order to ensure that only stable parts of the edge are considered, only pixels conserved in at least 9 out of 10 consecutive images are kept in the final picture. These final pictures, exemplified in Fig.\ref{fig:LEFTRIGHTFINAL}(a), were analyzed to extract the properties of the deposit. 
 
 Deposits are characterized by their spatial extension, conventionally called height $h(x,t)$, measured as the distance between the center line and the farthest white pixel along $y$, at a given $x$ (see Fig. \ref{fig:LEFTRIGHTFINAL}(b)). The height is initially computed independently on the left and right sides of the center line. The growth of the Marangoni edge is afterwards described through the relationship between $\bar h(t)$, the average of $h(x,t)$ along $y$, and $w(t)$, the standard deviation of $h(x,t)$, conventionally called width of the deposit (see Fig.\ref{fig:LEFTRIGHTFINAL}(b) for graphical definitions).

\subsection{Edge growth statistics}
\label{ScalingTheory}
Usually, in order to describe the deposit growth quantitatively, both its mean height $\bar h(t)$ and roughness width $w(L,t)$, which is the standard deviation of surface fluctuations around its mean value, are studied. For a discrete system, as our experimental pictures made of pixels, these two quantities are defined as \cite{barabasi_stanley_1995, FAMILY1990561}:
\begin{equation}\label{1}
	\bar h(t) = \frac{1}{L}\sum\limits_{i = 1}^L {h(i,t)}
\end{equation}
where $h(i,t)$ indicates the height of the deposit at the $\mathrm{i}$-th pixel along $x$ at the time $t$, and $L$ is the size of the system in pixels (see Fig.\ref{fig:LEFTRIGHTFINAL}(a)). If the deposition rate (the number of particles reaching the surface per unit of time) is constant, the mean height increases linearly with time ($\bar h \sim t$). To quantify the growth process, the roughness width is investigated as a function of time
\begin{equation}
	w(L,t) = \sqrt {\frac{1}{L}{\sum\limits_{i=1}^L\left[h(i,t)-\bar{h}(t)\right]^2}}\,. \label{2} 
\end{equation}
For many phenomena \cite{meakin1987restructuring,csahok1992kinetic,kurnaz1993sedimentation,kurnaz1996sedimentation,oliveira2011roughness,mccloud1997deposition,mccloud2002effect,cardak2006experimental}, the width $w$ increases as a power of time
\begin{equation}
	w(L,t)\propto t^{\beta}~,
		\label{eq:BetaEq}
\end{equation}
where $\beta$ is called the growth exponent and determines the time-dependent dynamics of the growth process. In experimental contexts, since defining the time zero is often difficult or arbitrary to some extent, a common approach consist in assuming that the scaling between $w$ and $t$ also translates between $w$ and $\bar h$: $	w(L,\bar h)\sim L^{\alpha}f(\frac{\bar h}{L^{z}})$.  In our experiments, this beta coefficient was then obtained by fitting an piece-wise function, defined as an arbitrary straight line, continuously connected to a horizontal line (see Fig.\ref{fig:LEFTRIGHTFINAL}(b)). The slope of the first line was considered to be the $\beta$ parameter. The final horizontal line was used in the fit, because the width eventually saturates at a value that increases as a power law of the system size:
\begin{equation}
	w_{sat}\propto L^{\alpha}
	\label{eq:AlphaEq} 
\end{equation}
The exponent $\alpha$ is the roughness exponent of the saturated surface.
The above relations are often expressed in the relation introduced by Family–Vicsek \cite{family1985scaling}:
\begin{equation}
	w(L,t)\propto L^{\alpha}f\left(\frac{t}{L^{z}}\right),
\end{equation}
where $z=\frac{\alpha}{\beta}$ is the dynamical exponent, and $f(u)\propto u^\beta$ if $u\ll1$ and approaches a constant when $u\gg1$ \cite{FAMILY1990561, family1985scaling}.

In order to measure the roughness coefficient $\alpha$ from Eq.(\ref{eq:AlphaEq}), we calculate the local width by dividing the pictures into windows of length $l$ along the deposit axis, and average the results obtained from all the windows \cite{sardari2024dynamics}, 
\begin{equation}
	w^2(l,t) = \left\langle\frac{1}{l}\sum_{x=x_0}^{x_0+l}[h(x,t)-\bar{h_l}(t)]^2\right\rangle_{x_0} \label{9} 
\end{equation}
where $ \bar{h_l}(t)$ is the average height of the selected window and $<>$ indicates the ensemble averaging. According to Eq.(\ref{eq:AlphaEq}), the local width increases exponentially as the window size increases. Experimentally, it has already been shown that this exponent can follow various regimes, with transitions at given system sizes $L$ \cite{mccloud2002effect}. The roughness exponent was then obtained by fitting the data obtained after $w(L,t)$ saturated with a continuous piece-wise function, with two successive lines having positive slopes, followed by a constant value. The first line's slope was considered as the roughness exponent at short-length scales $\alpha_s$. The slope of the second line is the roughness exponent at long length scales $\alpha_l$ (see Appendix Fig. \ref{fig:SuppDataAlpha} for an example of the fitting process).

The previous scaling relationships can be used to define universality classes. In the context of evaporating droplets, it has been previously shown that the exponent $\beta$ is sensitive to the shape of the particles, which can modify the growth regime of the edge formed by the coffee ring process. In particular, spheres have shown to follow a Poisson process, characterized by $\beta=0.5$ and $ \alpha\approx 0$, while  anisotropic particles, with anisotropic interactions, exhibited a growth belonging to the Kardar-Parisi-Zhang process with quenched disorder (KPZQ) with $\beta=0.68$ and $\alpha=0.63$ \cite{yunker2013effects}. Given that the RBC shape can be altered by physio-chemical stresses or in genetic diseases, one of the objective of this work is to determine if the growth of the observed Marangoni edges can belong to various growth mechanisms. In particular, we tested if artificial changes of RBCs shapes and aggregability can lead to a different growth class of the central Marangoni edge. To that end, we analyzed the growth of the Marangoni edge obtained with various suspensions, with the image analysis described previously.


\section{Results and discussion}

In the beginning, dried deposits were obtained from spherical cap droplets with different suspending liquids
(see Fig.\ref{fig:fig1}(a-c)): autologous plasma, $4.5\,\mathrm{mg/mL}$  Bovine Serum Albumine (BSA) (Sigma Aldrich, Burlington, Massachusetts, United States) solution in PBS, and in $1\,\mathrm{mL}$ of fibrinogen solutions (concentrations upto $5\,\mathrm{mg/mL}$ were used for fibrinogen in PBS, $4.5\,\mathrm{mg/mL}$ of BSA was also added in order to prevent the glass effect and the formation of slat crystals in the final dried deposits). As schematized in Fig.\ref{fig:figsetup}(c) and displayed in Fig.\ref{fig:fig1}(a-c), RBCs migrate and accumulate towards the center of the drop due to a dominating Marangoni flow. Hence, the pattern of all the dried drops in the conducted experiments includes central deposits of RBCs due to a Marangoni recirculation loop. In order to study the growth of this central deposit, linear droplets were placed on a glass slide, as illustrated in Fig.\ref{fig:fig1}(d-f). This made it possible to use classical 2D edge growth description, previously used to model coffee rings \cite{yunker2013effects} and described in the section \ref{ScalingTheory}. The dynamics of deposition was recorded at one frame per second with an 4X objective, around the middle of the droplet's line. A typical picture is displayed in the insert of Fig.\ref{fig:fig1}(d).

  \begin{figure}[ht!]
	\begin{center}
		\includegraphics[scale=0.390]{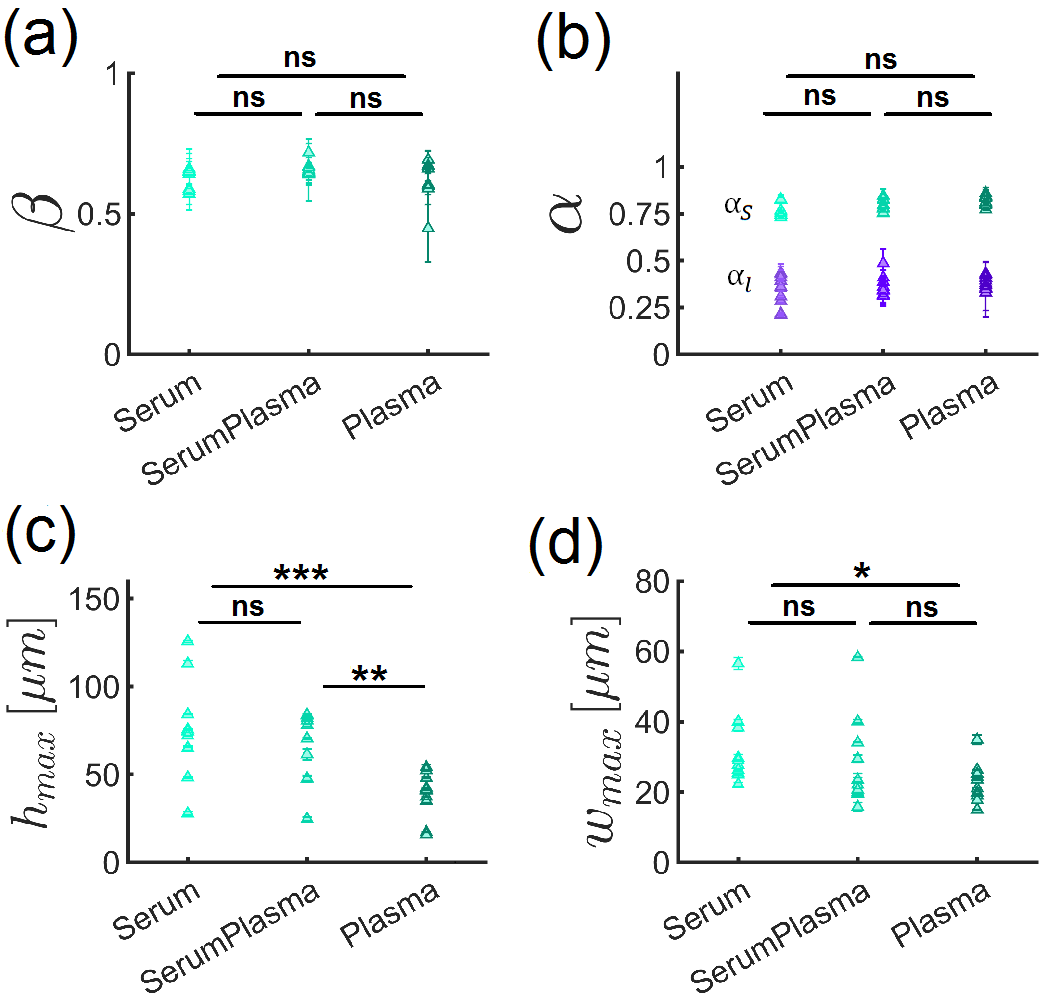}
		\caption{Experimental results: (a) Average growth exponents (b) Average roughness exponent (c) Maximum height (d) Maximum width, obtained for $0.3\%$ red blood cell suspension were formed in $1\,\mathrm{mL}$ Serum, Plasma and 50-50 mixture of autologous serum and plasma (SerumPlasma). Measurement has been repeated ten times for all three colloidal suspensions. The significance of the difference in the distributions is abbreviated as ns (not significant) for $p >0.1$, $*$ refers to a significance level of $p<0.1$, $**$  to  $p<0.01$ , and $***$ to $p<0.001$.
		}
		\label{fig:FIG1}
	\end{center}
\end{figure}

The main objective of this study is to determine whether evaporative deposits of dilute RBC suspensions could be used as low-cost screening methods for pathological modifications of RBC aggregability and shapes. In order to determine whether the aggregability of the RBC modifies the growth of the central deposit, we performed experiments by suspending 0.3\% volume of RBCs in plasma, serum and a 50-50 mixture of autologous serum and plasma.

Indeed, serum can basically be described as fibrinogen-free plasma, and fibrinogen is one of the main molecules promoting aggregation in blood samples. Using autologous serum then offers a practical way to modify the RBCs aggregability while maintaining other parameters as close as possible to their physiological conditions, as already employed in different studies \cite{dasanna2022erythrocyte,darras2023erythrocyte,john2024early}. The parameters obtained for three fibrinogen levels are displayed in Fig. \ref{fig:FIG1}, for fresh blood samples from 3 healthy donors and a repetition of 10 droplets. Although all suspension demonstrated a growth with an exponent close to the one of a KPZQ process ($\beta\approx 0.68$ for the KPZQ, our experimental values lie within $\beta\approx 0.64\pm0.04$), it is interesting to note that there is a significant difference in the distribution of the maximal roughness $w(L,t)$ and maximal average height $\bar h(t)$ reached by the deposits over the run of the experiment. These maximal values are typically reached at the end of the experiment, but noise and minor perturbation (such as irregularities in the shape of the droplet) can slightly shift the observation time of these values.
The significance of those variations is summarized by the range of p-values obtained for two-sample Kolmogorov-Smirnov tests. These ranges are abbreviated as ns (not significant) for $p > 0.1$, while the number $n$ of stars $*$ refers to a significance level of $p<10^{-n}$. Interestingly, the changes of $h_{max}$ and $w_{max}$ seem to be more sensitive to an increase of fibrinogen levels, meaning that one could potentially detect inflammation with a method based on this observation, as fibrinogen is a well-known inflammation marker \cite{germolec2018markers}. We can rule out that this change is due to a variation in the intensity of the Marangoni flow, because there is no significant variation of the surface tension between serum, plasma and serum-plasma mixture (see Appendix Fig.\ref{fig:SuppDataTension} for measurements with the pending drop method). The most obvious assumption is then that the modification of the RBC aggregation is the parameter responsible for this change \cite{brust2014plasma}.

Since RBCs are one of the most deformable cell types, whose shape can be altered in pathologies, we also performed experiments to investigate whether artificial changes in RBC shapes and aggregation can lead to variations in the growth of the central Marangoni edge. More accurately, we tested if shapes from the Stomatocyte-Discocyte-Echinocyte (SDE) sequence, obtained via a change of osmolarity, would lead to any difference in the dynamics of the deposit formation. Although the osmolarity of the suspending phase should increase during evaporation of the droplet, and possibly alter the shape of the RBCs during the experiments, we noticed that the significant part of the central deposit growth occurs within ten minutes of the evaporation process, while the overall evaporation process took around 40 minutes. The overall increase of the osmolarity over the formation of the central deposit is therefore around $1/0.75-1$, i.e., an increase of approximately $33\%$, assuming a constant evaporation rate.  This overall increase is too small to significantly alter the shape of the RBCs. Moreover the osmolarity is probably inhomogenous in the droplet, and the increase of osmolarity is probably limited to the edge of the droplets, since strong Marangoni recirculation is linked to a salt concentration gradient towards the edge of the droplet \cite{marin2019solutal,shahidzadeh2008salt,shahidzadeh2015salt}. Finally, a closer inspection of the dried patterns shows that discocytes are still forming rouleaux in the case of initial osmolarity of 300 mOsm/L and that the spherocytic or echinocytic shapes are still conserved in the dried deposit. Change in RBC chapes during the formation of the deposit is therefore negligible.

\begin{figure}[ht!]
	\begin{center}
	\textbf{$\phi=3\times 10^{-3}$}\par\medskip
		\includegraphics[scale=0.42]{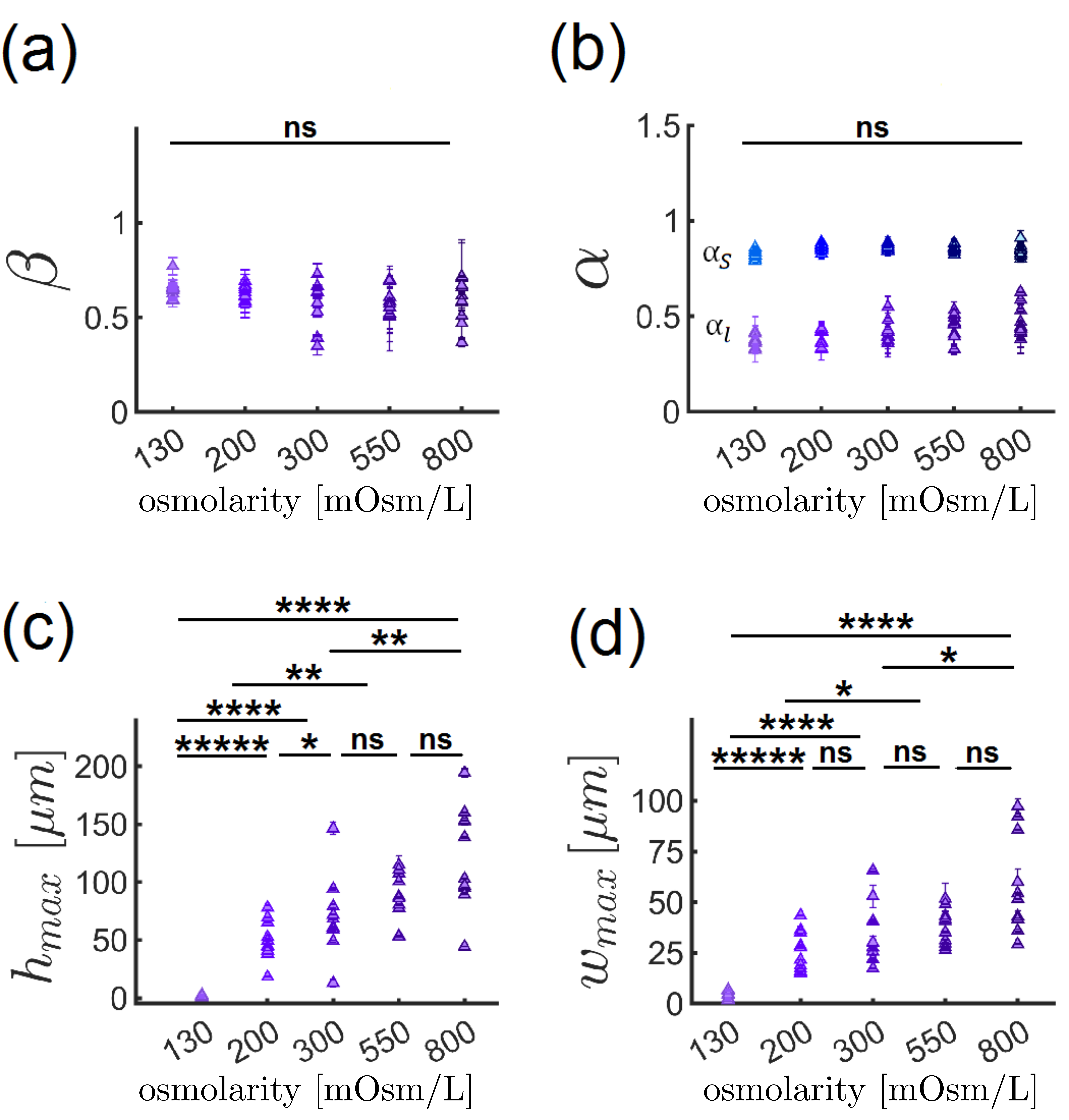}
		\caption{Experimental results: (a) Average growth exponents (b) Average roughness exponent (c) Maximum height (d) Maximum width, obtained for $0.3\%$ red blood cell suspension were formed in $1\,\mathrm{mL}$ solutions in osmolarity $(130, 200, 300, 550, 800)$ [ mOsmol/L]. Measurement has been repeated ten times for all three colloidal suspensions. The abbreviation ns stands for not significant. (The significance of p-values is defined as ns: $p >0.1$, $*$: $p<0.1$, $**$: $p<0.01$ , $***$:  $p<0.001$, and $*****$: $p<0.00001$).
	}
		\label{fig:FIG2}
	\end{center}
\end{figure}

\begin{figure}[ht!]
	\begin{center}
     \textbf{$\phi=3\times 10^{-3} \times V_{Osm}/V_{300}\Rightarrow N_{RBC}\approx const.$}\par\medskip
		\includegraphics[scale=0.42]{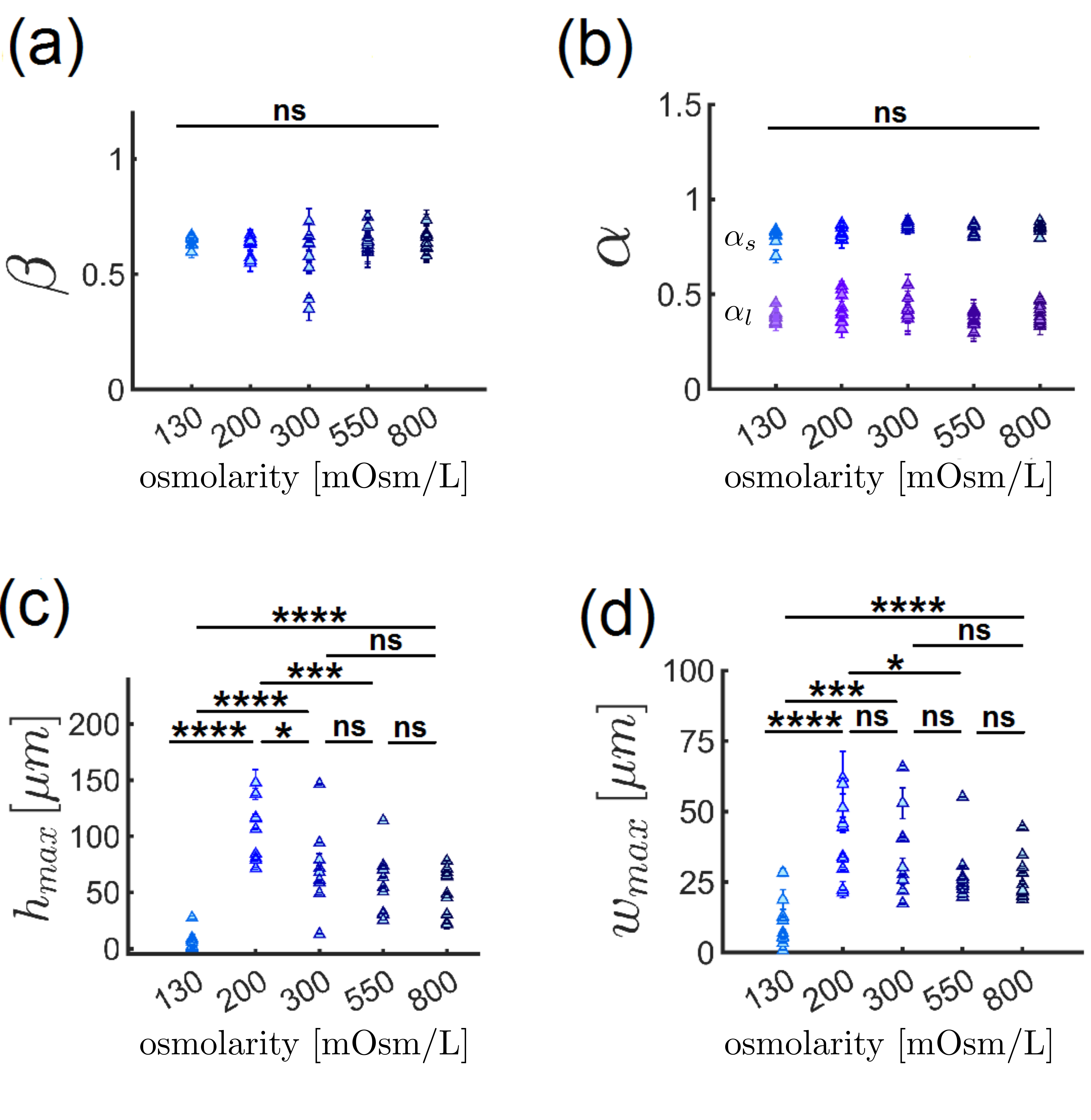}
		\caption{Experimental results: (a) Average growth exponents (b) Average roughness exponent (c) Maximum height (d) Maximum width, obtained for $0.3\%$ red blood cell suspension were formed in $1\,\mathrm{mL}$ solutions in osmolarity  with the equal number of blood cells per milliliter. Measurement has been repeated ten times for all three colloidal suspensions. The abbreviation ns stands for not significant. The significance of p-values is abbreviated as ns (not significant) for $p >0.1$, $*$ refers to a significance level of $p<0.1$, $**$  to  $p<0.01$ , $***$ to $p<0.001$, and $****$ to $p<0.0001$ .
}
		\label{fig:FIG3}
	\end{center}
\end{figure}
Growth, roughness exponent, maximum height and maximum width obtained for various osmolarity levels are displayed in Fig \ref{fig:FIG2}. According to previous results with inorganic particles, we were expecting some changes in the growth regime of the central deposit for various particles shapes, especially when moving form isotropic to anisotropic shapes \cite{yunker2011suppression,yunker2013effects}. The change in cell shape did not significantly impact the overall morphology or the universal class of growth. All samples were consistent with the KPZQ universality class, regardless of cell shape alterations. However, we also observed a significant increase of the maximal height of the central deposit, particularly sensitive to the change of osmolarity in the lower range, i.e., in the spherocyte-discocyte part of the SDE sequence. It is important to notice that the volume of each blood cell also exhibits a clear dependence on osmolarity \cite{friebel2010influence,heubusch1985osmotic} (see Appendix Fig. \ref{fig:SuppDataVolume}). Consequently, it was not clear at this stage whether this dependency is based on the shape of the erythrocytes or their volume in the droplets of various initial osmolarities. We could however rule out a change in the intensity of the Marangoni flow, as there was no significant difference in the surface tension of the various suspensions (see Appendix Fig. \ref{fig:SuppDataTension}).

We therefore performed an additional set of experiments, maintaining a constant number of blood cells, instead of maintaining a constant volume fraction of $\phi=3\times 10^{-3}$ across all osmolarities. To that end, we estimated the average volume of RBCs at various osmolarities $V_{Osm}$ through statistics of 3D scans performed with a confocal microscope (see Appendix Fig.\ref{fig:SuppDataVolume} for values), and suspended the RBCs with a volume fraction computed as $\phi=3\times 10^{-3} \times  V_{Osm}/V_{300mOsM}$, where $V_{300mOsM}=90\,\mathrm{\upmu m^3}$ is the standard volume of discocytes at physiological osmolarity of $300\,\mathrm{mOsmol/L}$ (compatible with our measurements) \cite{brugnara1987cell}. With this change, we maintained a comparable number of RBCs in all samples, rather than a similar volume. Results with this protocol are depicted in Fig.\ref{fig:FIG3}. Interestingly, a significant increase of the maximal height is still observed from $130\,\mathrm{mOsmol/L}$ to $200\,\mathrm{mOsmol/L}$, but the maximal width decreases or remains constant for higher osmolarity. We can then rule out that the osmolarity of the sample alone is responsible for the observed trend, as it is not reproduced if we change another parameter. Normalizing the dimensions of the central deposit by the length size of the cells (i.e., the cubic root of the average volume $V_{Osm}^{1/3}$ measure for a specific osmolarity), does not significantly alter the observed trends. Therefore, the volume of the cells alone is also not explaining the trends observed at low osmolarity in Figs.\ref{fig:FIG3} and \ref{fig:FIG2}, although the trend observed in Fig.\ref{fig:FIG3} at high osmolarity might be related to this change of volume, as correcting for the number of cells show that there is no significant differences in the height of the deposit beyond 300 mOsM in Fig. \ref{fig:FIG2} (see Appendix, Fig. \ref{fig:FIG2Ap} for the graph with normalization by the characteristic radius).\\

\section{Conclusion}

Our results showed that suspensions of RBCs at low volume fractions ($\phi=3\times 10^{-3}$) grow a central deposit, which contradicts the common assumption that blood evaporation is dominated by the classical coffee-ring effect \cite{pal2020concentration,zeid2013influence,chen2016blood,pal2021temperature,pal2023drying,sikarwar2016automatic,bahmani2017study,liu2020automatic,attinger2022using,roy2024insights}. These results already mean that the Marangoni stresses should be incorporated in models used to explain the apparition of cracks and patterns in dried blood deposits \cite{roy2024insights,hennessy2022drying}. We demonstrated that the growth of this central deposit follows a dynamics close to predictions of the KPZQ scalings. This KPZQ scaling is robust enough to be observed in all osmolarities where the SDE transition can be observed, independently of the shape adopted by the RBCs during the growth of the Marangoni edge. Our investigations also demonstrated that the maximal height (in terms of average spatial extension of the 2D deposit) of this edge is however depending on the properties of the RBCs. In particular, a higher aggregation of RBCs induces a smaller deposit, while spherocytes tend to create smaller deposits, with the highest degree of significance. These results indicate that the characterization of the size of this central deposit has potential application as a clinical tool to detect a change in cell volume and/or geometry. However, our results also show that this size has a smaller sensitivity to changes corresponding to higher osmolarity levels. This indicates that an optimization process would first be required fo practical applications, either by developing automatic and highly reproducible elongated droplet shapes, or by exploring the effect of other control parameters such as volume fraction, surrounding humidity and temperature. 

\section{Author contributions}
V.S. and A.D. designed the initial experimental methodology and wrote the image analysis software. V.S and M.M. developed the formal analysis. V.S. performed experiments and analyzed the results. A.D. obtained the ethical approval for the research. A.D. and C.W. obtained core funding for the research. L.K. and T.J., F.M., A.D. and D.\"O. developed the cell volume measurement technique with the confocal microscope,  D.\"O.performed the experimental measurements, analyzed by F. M. and A.D. R S. provided facilities for the surface tension measurements, performed by M.M., S. A. and V.S. V.S. and A.D. wrote the original draft of the manuscript. All authors reviewed and rewrote the manuscript.

\section{Ackowledgements}
V.S. ackowledges funding from Ministry of Science, Research, and Technology of Iran for her research stay in Germany.  A.D. acknowledges funding by the Young Investigator Grant of the Saarland University.

\bibliographystyle{elsarticle-num}
\bibliography{BiblioBloodDrop}

\setcounter{figure}{0} 
\appendix
\section{Supporting information}

\begin{figure}[ht!]
	\begin{center}
		\caption{(A) The final shape of the dried droplets after the appearance of cracks and salt crystals in the various suspensions with initial RBC volume fraction of $0.3\%$. Supernatants were (a and d) autologous plasma, (b and e) $4.5\,\mathrm{mg/mL}$  BSA solution in PBS, and (c and f) $5\,\mathrm{mg/mL}$ fibrinogen solution in PBS with $4.5\,\mathrm{mg/mL}$  BSA and (g) Serum. The final shape of the dried drops despite the crack pattern and crystals. Center line in drops of dried blood, (B) with equal volume of blood cells in each sample, (C) with equal number of blood cells in each sample of various osmolarity. See Appendix Fig. \ref{fig:SuppDataVolume} for volume measurements.}
		\includegraphics[scale=0.28]{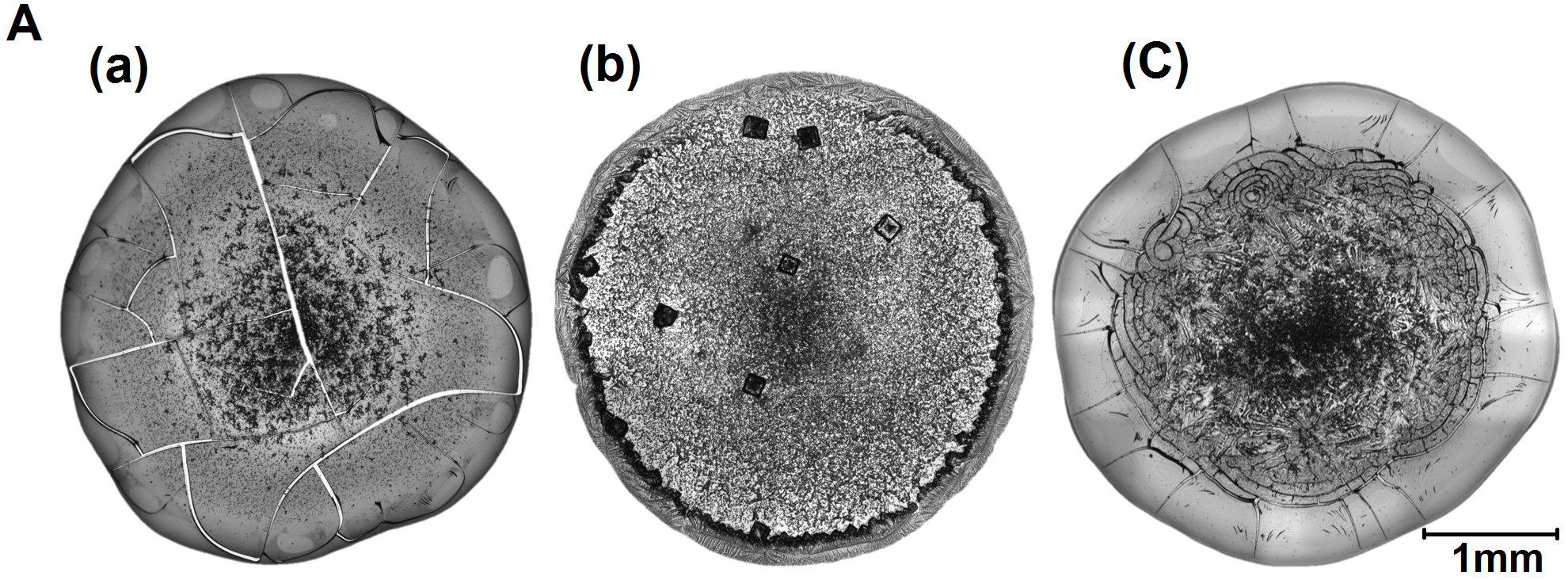}
				\includegraphics[scale=0.28]{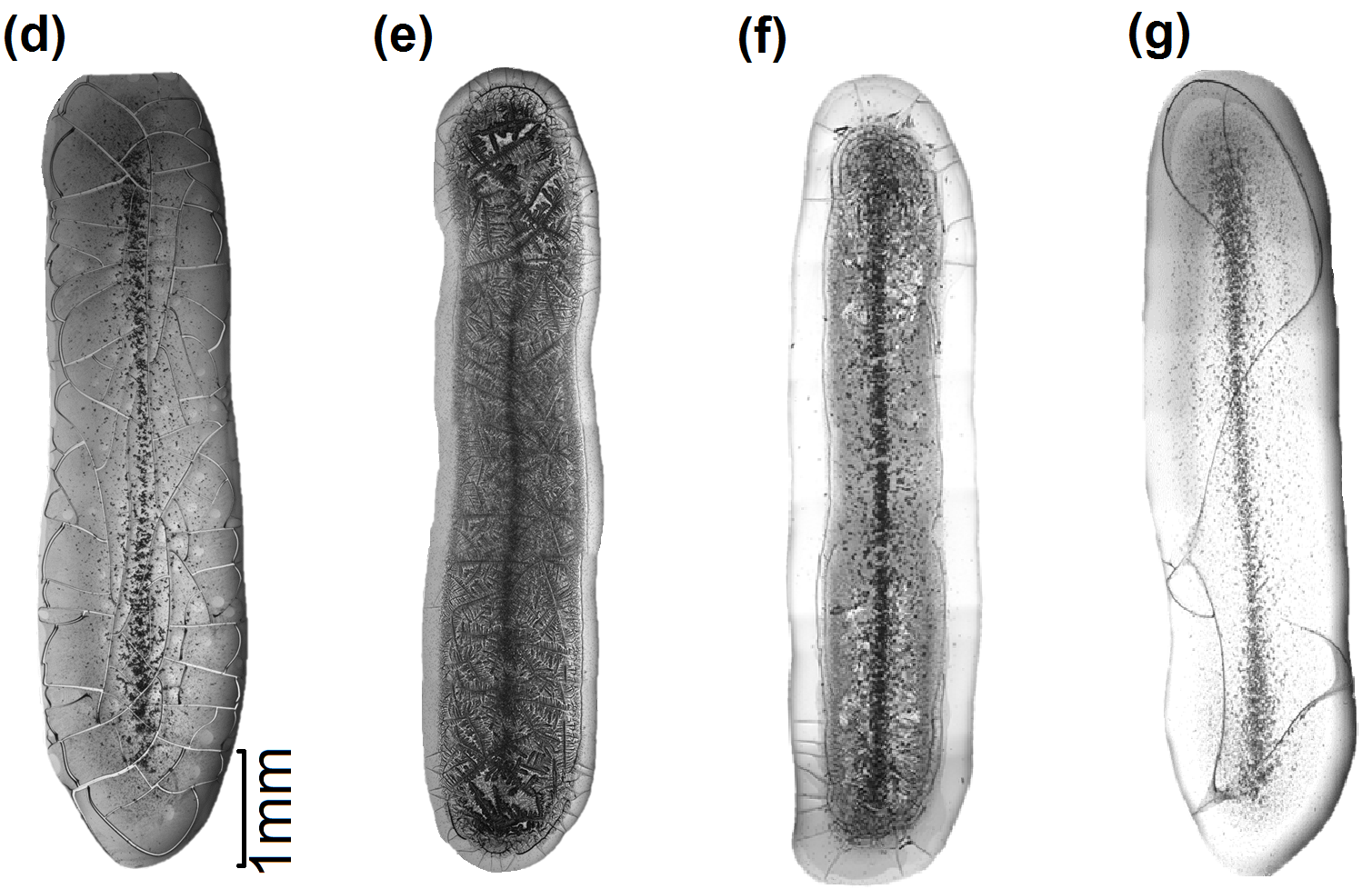}
			\includegraphics[scale=0.37]{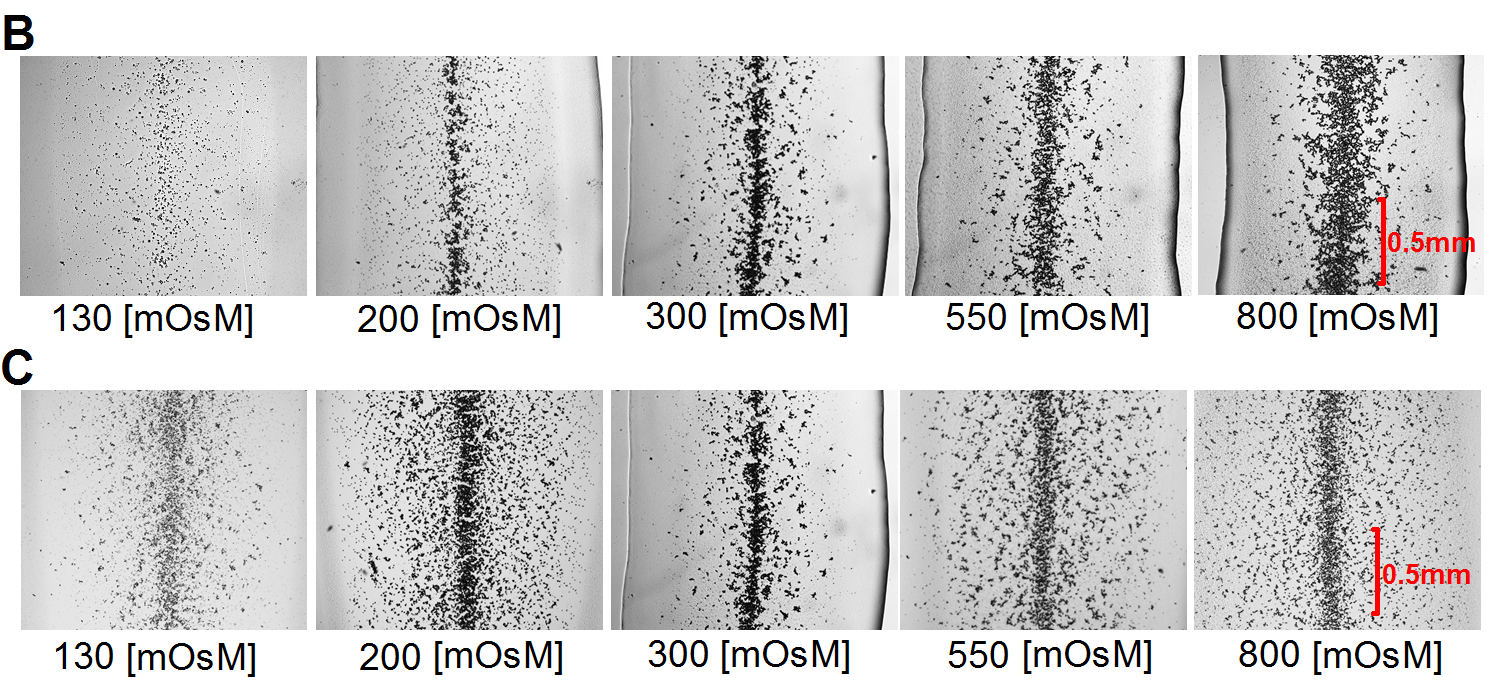}
		\label{fig:SuppDeposits}
	\end{center}
\end{figure}

\begin{figure}[ht!]
	\begin{center}
		\includegraphics[width=0.85\textwidth]{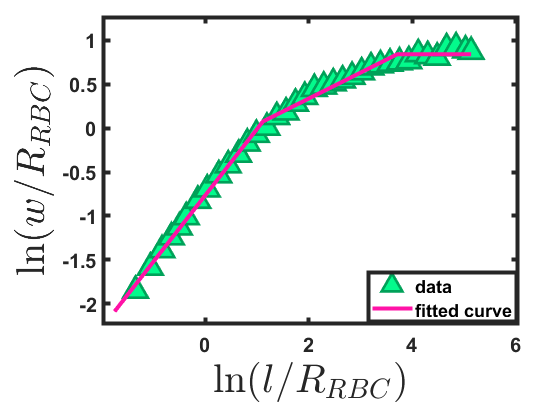}
		\caption{Scaling of the interface width with the size of the measurement window. A continuous piece-wise function, with two lines followed by a constant value, is fitted to the data in order to determine two $\alpha$ coefficients, $\alpha_s$ and $\alpha_l$ from the first and second slope, respectively \cite{mccloud2002effect}.}
		\label{fig:SuppDataAlpha}
	\end{center}
\end{figure}

\begin{figure}[ht!]
	\begin{center}
		\includegraphics[scale=0.60]{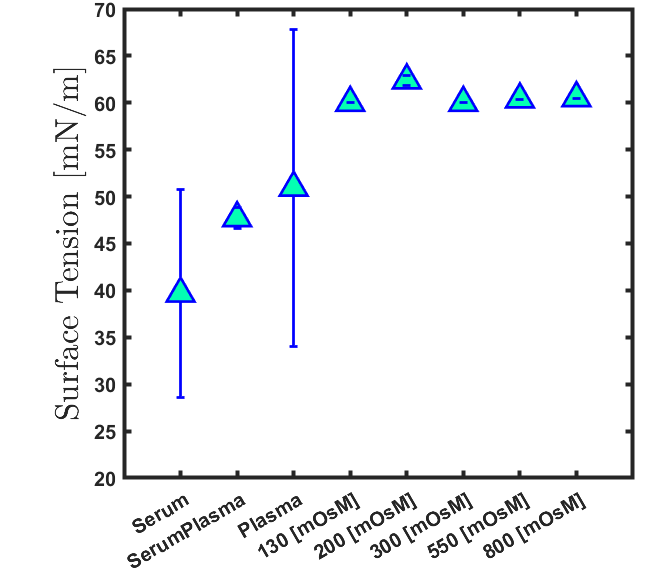}
		\caption{Liquid-air surface tension of the various liquid phases used in this study, measured by the pending drop method. While the ionic buffer has a surface tension higher than the serum of various donors, the changes of osmolarities probed in this work did not induce significant changes. A higher dispersion of surface tension is obtained for the various donors, although plasma, serum and serum-plasma mixtures do not have significantly different surface tension, nor consistent trend for a given donor.}
		\label{fig:SuppDataTension}
	\end{center}
\end{figure}

\begin{figure}[ht!]
	\begin{center}
		\includegraphics[scale=0.55]{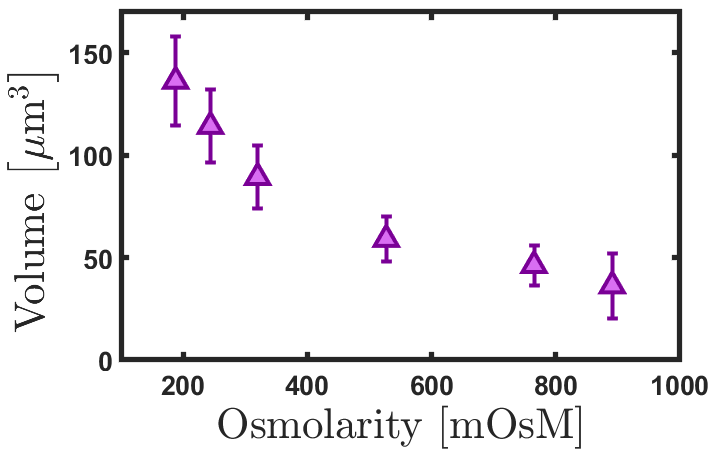}\\
		\caption{Volume of RBCs measured at various osmolarities, for 4 healthy donors. Measurements were performed on a fresh cell population, obtained from finger prick. The cells have been washed, stained with CellMask Deep Red, then washed in the solution with aimed osmolarity. Suspensions with 0.5\% volume fraction of RBCs are then 3D scanned with a confocal microscope. Values are obtained after isolating the scan of each individual cell and removing outliers. Values are in agreement with data available in the literature for other buffers \cite{friebel2010influence,heubusch1985osmotic}.}
		\label{fig:SuppDataVolume}
	\end{center}
\end{figure}
\begin{figure}[ht!]
	\begin{center}
	\textbf{$\phi=3\times 10^{-3}$}\par\medskip
		\includegraphics[scale=0.22]{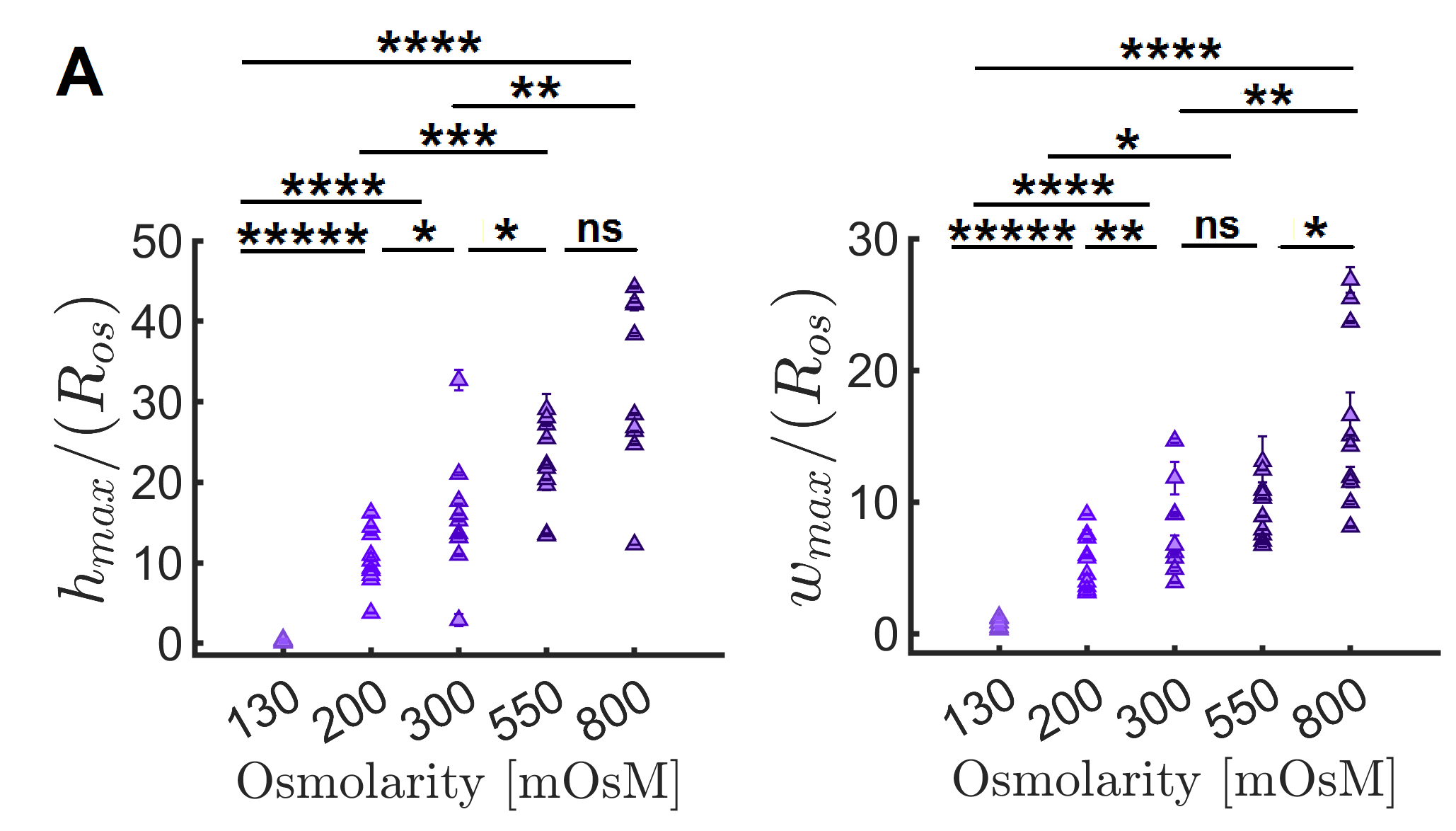}
		
		\includegraphics[scale=0.22]{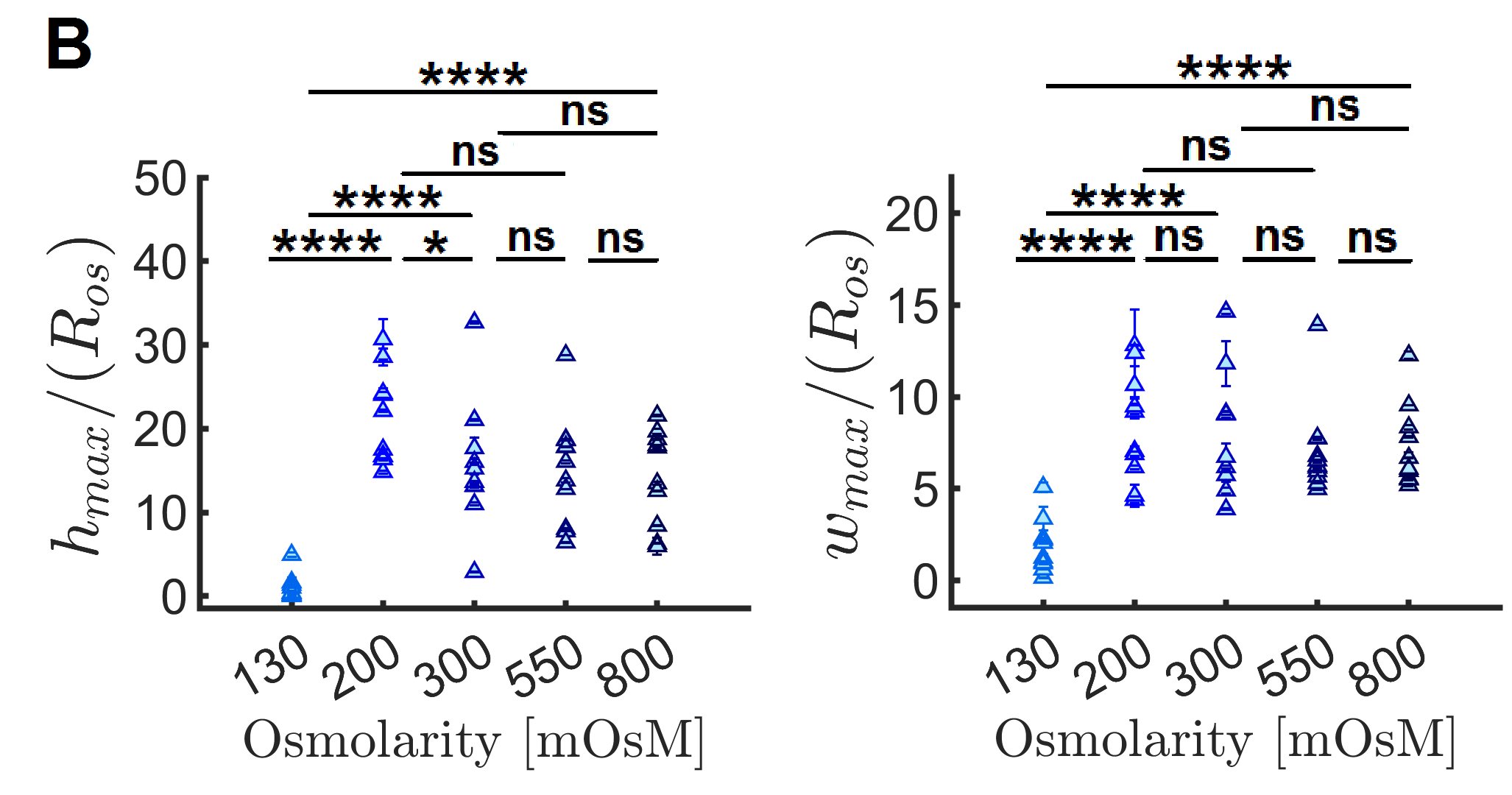}
		    \textbf{$\phi=3\times 10^{-3} \times V_{Osm}/V_{300}\Rightarrow N_{RBC}\approx const.$}\par\medskip
		\caption{Experimental results: (A) Maximum height and width normalized by cell radius at each osmolarity, obtained for $0.3\%$ red blood cell suspension were formed in $1\,\mathrm{mL}$ solutions of various osmolarities. (B) Maximum height and width normalized by cell radius at each osmolarity, obtained for $0.3\%$ red blood cell suspension were formed in $1\,\mathrm{mL}$ solutions of various osmolarities with an equal number of blood cells per milliliter. Measurement has been repeated ten times for all three colloidal suspensions. The abbreviation ns stands for not significant. (The significance of p-values is defined as ns:$p >0.1$, $*$:$p<0.1$, $**$:$p<0.01$ , $***$:$p<0.001$, and $*****$:$p<0.00001$).
	}
		\label{fig:FIG2Ap}
	\end{center}
\end{figure}






\end{document}